# Towards Secure and Reliable Heterogeneous Real-time Telemetry Communication in Autonomous UAV Swarms

_Pavlo Mykytyn_[1,2], Marcin Brzozowski[1], Zoya Dyka[1,2], Peter Langendörfer[1,2]
[1] IHP - Leibniz-Institut für innovative Mikroelektronik, Frankfurt (Oder), Germany,
[2] BTU Cottbus-Senftenberg, Cottbus, Germany
pavlo.mykytyn@b-tu.de

**Summary:**
In the era of cutting-edge autonomous systems, Unmanned Aerial Vehicles (UAVs) are becoming an essential part of the solutions for numerous complex challenges. This paper evaluates UAV peer-to-peer telemetry communication, highlighting its security vulnerabilities and explores a transition to a heterogeneous multi-hop mesh all-to-all communication architecture to increase inter-swarm connectivity and reliability. Additionally, we suggest a symmetric key agreement and data encryption mechanism implementation for inter - swarm communication, to ensure data integrity and confidentiality without compromising performance.

## 1. Introduction

The increasing integration of UAVs into various commercial and private solutions underscores the critical need for robust communication protocols ensuring the confidentiality, integrity, and availability of transmittable data. Our objective is to address the challenges associated with secure and reliable point-to-point, point-to-multipoint and all-to-all telemetry communication within UAV swarms and describe the lessons-learned along the way. Autonomous UAVs, as well as UAV swarms, require a radio telemetry link to establish a connection with the ground control station (GCS). This connection allows remote control through operator input or autonomous operation based on a preprogrammed mission. To facilitate and organize the communication between a GCS and airborne or terrestrial robots, Lorenz Meier introduced the Micro Air Vehicle Link (MAVlink) communication protocol in 2009 under the LGPL license [1]. It describes a complete set of instructions that are sent forth and back between a UAV and a GCS. It is employed by major open source autopilot systems such as ArduPilot and PixHawk, and provides powerful features for monitoring and control of autonomous missions. The existing versions, v1.0 and v2.0 of the protocol, however, are lacking essential security measures and make the communication susceptible to eavesdropping, Man-In-The-Middle (MITM), replay, unauthorized access, and potential hijacking attacks. Notably, challenges related to MAVlink protocol encryption [2] further complicate matters and demand for other effective higher-level solutions that do not require modification of the protocol's source code. Striking a balance between efficient yet effective implementation of security measures is vital, considering the constrained computational resources of UAVs. Additionally, it is critical to properly address packet loss issues in wireless communication and make sure that additional security measures do not intensify them.

**Telemetry Radios**

To facilitate telemetry communication between a GCS and a UAV, the 433 MHz point-to-point small, light and inexpensive open source radio platform called SiK Telemetry Radios manufactured by 3DR and Holybro [3], running the bespoke open source SiK firmware are widely used in research projects. However, these radios have proven to be suitable only for pairwise communication and unsuitable for any swarm related solutions. They also do not provide any security features. In addition, by exploring other radio alternatives running modifications of SiK firmware in the Sub-1-GHz spectrum, we identified the RFD868x radios with multi-point and asynchronous non-forwarding mesh communication capabilities as well as hardware accelerated AES encryption support from RFDesign [4]. However, further research unveiled a more complex issue related to the legal regulations in the 800-900 MHz frequency band posed by many European countries, including Germany. Realizing a reliable implementation of asynchronous mesh communication architecture in the Sub-1-GHz band using the RFD868x or any other radios, for that matter, is extremely challenging due to the legal limitations. The 863-869 MHz band belongs to the ISM (Industrial, Scientific and Medical) category and is allowed for unlicensed use, however, in effect, only parts of the band are open for unlicensed use with additional limitations on transmission power and duty cycle [5]. The combination of legal constraints and limited spectrum availability encourages the employment of additional ISM frequencies, such as 2.4 GHz, for more reliable UAV inter-swarm communication.

## 2. Related Work

Several studies have been conducted on secure telemetry communication in UAVs. However, to the best of our knowledge, there are no studies that simultaneously cover communication and routing protocols, suitable radio communication technologies, heterogeneous real-time inter-swarm communication approaches and suggest data encryption or authentication mechanisms. Nevertheless, some related works have studied parts of these aspects independently. The authors in [6], for instance, performed an empirical analysis study of MAVlink protocol vulnerabilities, demonstrating an attack method leveraging the unencrypted communication to disable a UAV. Through Internet Control Message Protocol flooding and packet injection attack, they discovered that it is possible to stop the mission, delete mission information and even take full control of the UAV. The authors in [7] discussed two different message propagation techniques in swarm communication, namely message routing and message flooding. In a routing-based approach, messages follow a designated path from node to node until the destination is reached. Contrary to routing, the flooding technique is based on broadcasting messages to all nodes in the network simultaneously. It simplifies network management by eliminating the need for routing, self-discovery, and repair algorithms, but requires extra resources. The authors stated that routing approaches consume less energy in smaller networks with few hops, however in larger networks or when message size is small, flooding eliminates the overhead associated with routing tables. Standard MAVlink command messages are much smaller than 256 bytes. Therefore, combined MAVlink messages could be composed into a single IEEE 802.11 frame during forwarding. Flooding-based networks, with mesh topology also provide lower latency due to lower overhead, which could be a critical factor for collision avoidance or other real-time tasks, while routing-based networks may experience inconsistent latency caused by RF propagation delay. To evaluate applications of routing approaches in Wireless Mesh Networks (WMNs) the authors in [8] conducted an analysis of well-established unsecured mesh and routing protocols such as Hybrid Wireless Mesh Protocol (HWMP), Better Approach To Mobile Ad-hoc Networking (BATMAN), and Optimized Link State Routing (OLSR). They proposed combining these routing protocols with the security frameworks of IEEE802.11s or IEEE802.11i standards. To properly evaluate the impact of these security frameworks on WMN operation they performed simulations and real testbed experiments. The experimental data have shown that black hole and wormhole attacks [9] were still feasible. A solution would require an efficient higher-lever security implementation combined with dynamic key management schemes. The authors in [10] introduced a Swarm Broadcast Protocol for dynamic leader-follower-based UAV swarms formations. Their approach is based on generating a broadcast key using conventional Diffie-Hellman key agreement in a chain sequence for each swarm. If a drone joins or leaves the swarm, a new sequence is created and a common broadcast key has to be regenerated based on the number of drones joining or leaving the swarm. To authenticate this exchange the authors mentioned two asymmetric sign-verify key pairs. The downside of this approach is the increased overhead related with each incremental broadcast key regeneration. Additionally, relying on the leader-drone for key generation introduces a single point of failure for the whole swarm, in case the leader-drone is unavailable. The rest of the drones in the swarm apart from the leader also require to be overlooked by an administrator during the signing of their "join request message" in the initialization phase. The mentioned studies have explored aspects of secure WMNs, communication protocols, and security vulnerabilities of UAVs. This study describes hands-on experience and challenges associated with developing and managing a dynamic UAV swarm and contributes to the development of a secure multi-hop all-to-all communication approach, offering a fresh perspective on inter-swarm communication strategies.

## 3. Transition to the Heterogeneous Approach

In contrast to the limitations and challenges associated with homogeneous telemetry communication, particularly evident in the Sub-GHz frequency band as mentioned in Section 1, a shift towards heterogeneous telemetry communication emerges as a much needed alternative. To effectively mitigate the effects of cyber-attacks and enhance reliability of inter-swarm and UAV to GCS communication, it is crucial to introduce diversification of the communication stack and distribute communication technologies across various operational frequencies, such as Sub-GHz, 2.4 GHz Wi-Fi and 5G/LTE. Our previous works [11, 12], describe the fundamentals for this approach and focus on development and integration of custom hardware components to enable heterogeneous telemetry communication.

**Point-to-point Communication**

Point-to-point topology is characterized by a direct communication link between a sender and a receiver, in our case, between a GCS and a UAV, allowing for data transmission with no involvement of intermediary nodes. While point-to-point communication is efficient, it is not suitable for swarm applications due to scalability issues. Swarm applications require more complex centralized or distributed architectures based on point-to-multipoint or mesh topologies to properly address the scalability requirements and minimize inter-UAV communication overhead. We have addressed this topic in greater detail in one of our previous works [13].

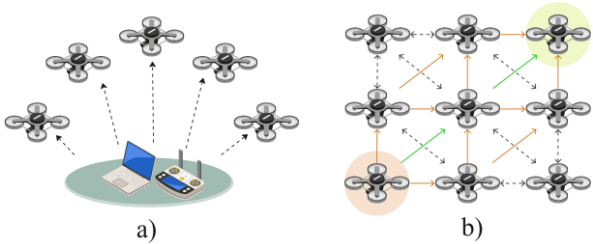

Fig. 1. a) Centralized star-shaped architecture; b) Decentralized multi-hop mesh architecture

**Point-to-multipoint and Mesh Communication**

Point-to-multipoint topology allows the transmitter for a simultaneous connection to multiple receivers, which is often implemented in a centralized, star-shape architecture depicted on Fig. 1a, in which one GCS can transmit and receive data from all of the UAVs in a swarm simultaneously. However, each UAV to UAV communication in this case still has to be routed through the GCS, which is not very efficient, although proven to be suitable for some application scenarios. The mesh topology represented on Fig. 1b, on the other hand, allows for a decentralized multi-hop all-to-all communication. In this case, all UAVs in a swarm that are within range can communicate directly or by utilizing other UAVs nearby to forward their messages when out of range.

**Hardware Requirements**

To diversify and distribute the communication over various frequencies, we chose the following radio modules. In the Sub-1-GHz band we picked Digi XBee SX 868 based on the Silabs EFM32 microcontroller capable of point-to-multipoint and DigiMesh [14] long range low throughput communication. In the 2.4 GHz band we selected Espressif ESP32 SoC with integrated Wi-Fi and Bluetooth LE for short range high throughput IEEE 802.11 communication. For the internet and cloud-based communication we chose SIMCom SIM7080G Cat-M module. During the design and development phase we observed that neither the PixHawk flight controller nor any suitable single-board companion computers such as Raspberry Pi or NVIDIA Jetson proved in having sufficient number of communication ports to connect and operate multiple radio modules simultaneously. Therefore, a custom-designed and built PCB to efficiently connect, manage, and operate multiple RF modules, each potentially utilizing different serial communication protocols such as SPI, I2C or UART becomes essential. We called this custom-developed PCB - Communication Hub and described its design and development process in our other work [11]. Additionally, a robust companion computer, such as Raspberry Pi 4B with 8 GB of RAM, becomes necessary for controlling and managing the Communication Hub and other collision avoidance related sensors. Based on the current network conditions, a companion computer should be able to seamlessly switch between different RF modules on the Communication Hub without significant packet loss, ensuring optimal performance and adaptability in dynamic environments.

## 4. Encryption Key Agreement and Authentication

To secure inter-swarm communication, we tailored a mechanism involving Elliptic Curves and both, symmetric and asymmetric cryptography. Our broadcast key agreement mechanism takes place over the air at the beginning of each flight mission. We define the UAV swarm as $\Sigma = \{\mu_1, \mu_2, \mu_3, \dots \mu_n\}$, where $\mu_n$ is the $n$ - th UAV. To agree on a session key, we utilize pairwise Elliptic Curve Diffie-Hellman (ECDH) key agreement between each $\mu_{1,2,\dots n}$ in $\Sigma$ and the GCS. The pairwise key agreement process is depicted on

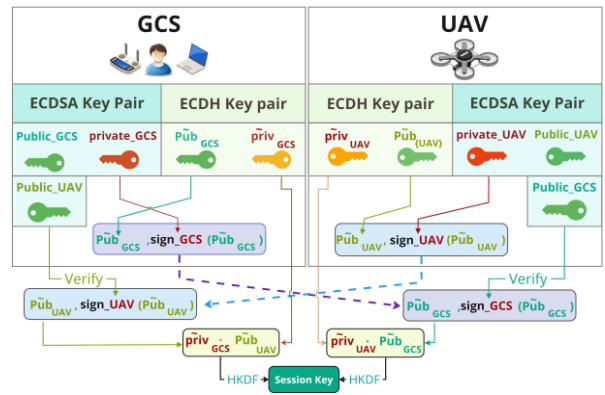

Fig. 2. Pairwise key agreement between GCS and UAV

Fig 2. However, ECDH alone does not prevent MITM attacks. Thus, to authenticate the key agreement process, we provide both GCS and $\mu_{1,2,...n}$ with their own set off Elliptic Curve Digital Signature Algorithm (ECDSA) keys beforehand. Additionally, the GCS's Public ECDSA key is preloaded into every UAV's flash memory. Correspondingly, each UAV's Public ECDSA key is stored in the GCS memory. Once the pairwise ECDH key agreement was initiated, GCS's public ECDH key is signed using its private ECDSA key and is distributed to each $\mu_{1,2,...n}$. Correspondingly, each $\mu_{1,2,...n}$ signs their public ECDH key with their ECDSA private key and sends it to the GCS. Upon receipt, ECDH public keys are verified for authenticity by each party, and then used to compute a common shared secret and generate a session key. Once the pairwise session key with each $\mu_{1,2,...n}$ has been established we can use it to symmetrically encrypt and communicate a time-out based rolling broadcast key to each $\mu_{1,2,...n}$ to ensure forward secrecy. After each $\mu_{1,2,...n}$ has received the first rolling broadcast key, all-to-all multi-hop mesh communication can be secured by using Advanced Encryption Standard (AES) in Galois-Counter Mode (GCM). AES-GCM is capable of high throughput high-speed communication. AES-GCM conducts Galois field multiplication of the cipher text with the hash of the plaintext and associated data to provide authenticity and integrity. The result of the multiplication is combined with encrypted message to form an authentication tag, which is then appended at the end of the cipher text.

## 5. Conclusion

To conclude, in this paper, we examined telemetry radio options, communication and routing protocols, and described current challenges and security vulnerabilities of secure and reliable inter-swarm telemetry communication. We portrayed our transition path to a heterogeneous multi-hop mesh all-to-all communication and described our lessons learned along the way. Additionally, we proposed an implementation of a key agreement and data encryption mechanism specifically tailored to accommodate all-to-all communication topology and ensure integrity, authenticity, and confidentiality of transmitted data without compromising performance.